\documentclass[fleqn,usenatbib,letters]{mnras}
\usepackage{newtxtext,newtxmath}

\usepackage[T1]{fontenc}
\usepackage{ae,aecompl}

\usepackage{graphicx}	
\usepackage{amsmath}	
\usepackage{amssymb}	
\usepackage{pifont}     

\def\aj{AJ}
\def\apj{ApJ}
\def\apjs{ApJS}
\def\apjl{ApJL}
\def\mnras{MNRAS}
\def\nat{\rm Nature}

\def\araa{ARA\&A}

\newcommand{\rapo}{R_{\rm apo}}
\newcommand{\dvtan}{\Delta v_{\rm tan}}
\newcommand{\tprime}{\theta^\prime}
\def\kms{ {\rm km~s\textsuperscript{-1}}}
\def\msun{ {\rm M_\odot}}
\newcommand{\mlmc}{M_{\rm LMC}}

\begin{document}
\label{firstpage}
\pagerange{\pageref{firstpage}--\pageref{lastpage}}

\title[MW Reflex Motion Signatures]
 {Reflex motion in the Milky Way stellar halo resulting from the Large Magellanic Cloud infall}

\author[Petersen \& Pe{\~n}arrubia] 
{Michael~S.~Petersen$^1$\thanks{michael.petersen@roe.ac.uk}
 \& Jorge~Pe{\~n}arrubia$^1$ \\ 
$^1$Institute for Astronomy, University of Edinburgh, Royal Observatory, Blackford Hill, Edinburgh EH9 3HJ, UK\\ }

\date{Accepted XXX. Received YYY; in original form ZZZ}

\pubyear{2020}

\maketitle

\begin{abstract}
We present the results of $N$-body models of the Milky Way and Large Magellanic Cloud system and study the kinematic reflex motion in the stellar halo owing to the barycentre displacement of the disc. In agreement with previous studies, we find that the Milky Way disc may be moving at 40~$\kms$ relative to the barycentre prior to the Large Magellanic Cloud infall. The resulting reflex motion is visible in tangential velocities of the stellar halo as a simple dipole. The signal is strongest for stars with long dynamical times, identifiable in position-velocity data as stars with large apocentres, whose dynamical memory is still well-represented by the unperturbed Milky Way potential. The signal varies across the sky depending on the stellar tracer and may be up to the same magnitude as the velocity of the disc centre-of-mass, making reflex motion a source of bias for Milky Way potential determinations based on the modeling of stellar streams and/or smooth halo tracers such as blue horizontal branch or RR Lyrae stars.
\end{abstract} 

\begin{keywords} 
galaxies: Galaxy: halo---galaxies: kinematics and dynamics
\end{keywords}

\section{Introduction} \label{sec:introduction} 

The recent treasure trove of kinematic data produced by the Gaia satellite \citep{gaia18} both enables and demands sophisticated modeling of the Milky Way (MW) potential to explain the observed kinematic features of stellar streams and smooth stellar halo tracers. One method to build model potentials of the MW is to fit orbits to stellar streams. Unfortunately, the results for the MW potential have been far from clear. Fits to the most prominent stream,  Sagittarius, have found best-fitting dark matter (DM) halo shapes that are strongly oblate \citep{law10}, or prolate \citep{johnston05,fardal19}. Fits to another prominent stream, GD-1, have found spherical \citep{bovy16} or oblate \citep{malhan19} DM halos.

Additionally, the presence of the Large Magellanic Cloud (LMC) challenges equilibrium models and prompts study of $N$-body MW models that include the LMC \citep{laporte18a,garavito19}. Studies suggest that the LMC is infalling for the first time with a mass between 1 and 3$\times10^{11}\msun$ \citep{besla12,kallivayalil13,penarrubia16,shao18,erkal19,wan19}. \citet{erkal19} used an improved hybrid technique to fit the Orphan stream \citep{koposov19} in the presence of the LMC: the MW DM halo and disc were allowed to move as a rigid bodies relative to the LMC, and the Orphan stream was fit in the resulting time-dependent potential. However, \citet{erkal19} found that the stream track could be fit with approximately equal quality using an oblate, spherical, or prolate DM halo. The fits produced significantly different parameters for the MW DM halo, including a 50 per cent variation in the virial mass and scale radius.

Furthermore, the kinematics of the smooth stellar halo are poorly understood. Studies have found evidence for mild rotation, the amplitude of which varies with the tracer: Blue Horizontal Branch (BHB) stars, RR Lyrae stars, and K giants all produced different values of $\langle v_\phi \rangle$ \citep{deason17}. \citet{kim19} also found evidence for change in orbital behaviour at $r\approx 30$ kpc that remains unexplained. 

To date, studies that model the dynamics of smooth halo tracers typically neglect the presence of the LMC. Yet,  \citet{gomez15} highlighted that the barycentre of the MW disc would be shifted as a response to the LMC infall.  The resultant reflex motion is an all-sky effect that may bias constraints on the MW potential when using individual stellar streams or tracers. In this paper, we study the reflex motion of the disc and the observable signature in the kinematics of the stellar halo. We build an illustrative $N$-body MW-LMC model and interpret the observed features in the stellar halo with a comprehensive suite of idealised tests designed to dismantle the complications of the reflex motion signal.

\section{Methods} \label{sec:methods}

We present live $N$-body models to test the LMC effect on MW motion and determine observable signatures.

\subsection{Model Components} \label{subsec:discandhalo}

{\bf Dark matter halo model.} The model is a spherically-symmetric Navarro-Frank-White (NFW) dark matter halo radial profile \citep{navarro97} given by $\rho_{\rm NFW}(r) \propto r_s^3r^{-1}\left(r+r_s\right)^{-2}$. The scale radius is set to be $r_s=0.04R_{\rm vir}$, where $R_{\rm vir}$ is the virial radius. We apply an error function truncation such that the initial halo profile is $\rho_h(r)=\frac{1}{2}\rho_{\rm NFW}(r)\left(1-{\rm erf}\left[(r-r_{\rm trunc})/w_{\rm trunc}\right]\right)$. The truncation parameters are $r_{\rm trunc}=2R_{\rm vir}$ and $w_{\rm trunc}=0.3R_{\rm vir}$. We realize the initial positions and velocities in the DM halo via Eddington inversion. The DM halo starts with no rotation and an isotropic distribution, described in \citet{petersen20a}. For the MW, a reasonable value for $R_{\rm vir}$ is 282 kpc \citep{blandhawthorn16}, making $r_s=$14.1 kpc. We choose the virial mass of the DM halo to be consistent with recent measurements of the MW, $M_{\rm vir}=M(<R_{\rm vir})=1.3\times10^{12}\msun$ \citep{blandhawthorn16}. The DM halo has $N_{\rm halo}=10^7$. We employ a `multimass' scheme for the DM halo to increase the number of particles in the vicinity of the disc. 

{\bf Stellar halo model.} The stellar halo is not a separate component, but rather a weighting of the DM halo particles. We weight each particle in the DM halo according to a chosen density profile to approximate the stellar halo, as in \citep{errani19}. We choose the stellar halo BHB density profile of \citet{deason14}. 

{\bf Disc model.} The stellar disc density is given by $\rho_d(r,z)~=~(M_{\rm d}/8\pi z_0R_d^2)~e^{-r/R_d} {\rm sech}^2~(z/z_0)$
where $M_d$ is the disc mass, $R_d=0.01R_{\rm vir}=2.8$ kpc is the disc scale length, and $z_0=0.001R_{\rm vir}=280$ pc is the disc scale height. We choose $M_d=0.025M_{\rm vir}=3\times10^{10}\msun$. We select the initial positions in the disc via an acceptance--rejection algorithm. We select the velocities by solving the Jeans equations in the disc plane, as in \citet{petersen20a}. We choose Toomre $Q=1.6$ to lessen disc instabilities during the simulation. The disc has $N_{\rm disc}=10^6$. 

{\bf LMC model.} The LMC model is a softened point source with a core of radius $0.04R_{\rm vir}=11.3$ kpc. We set 1:10 as the mass ratio with the DM halo, $M_{\rm LMC}=0.1M_{\rm vir}=1.2\times10^{11}\msun$.

\begin{figure} \centering \includegraphics[width=2.6in]{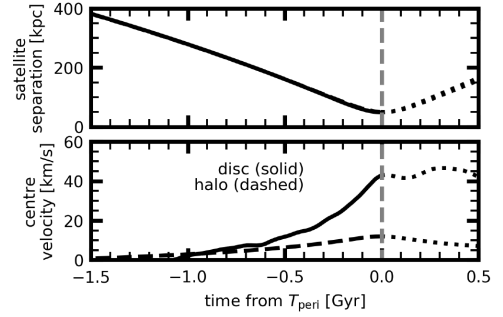} 
\caption{\label{fig:s13motion} Top panel: radial separation versus time for the disc and satellite centres. Bottom panel: centre of mass velocity relative to the satellite for the disc (solid) and DM halo (dashed). The dotted black curves indicate `future' evolution.} \end{figure}

\begin{figure*} \centering \includegraphics[width=6.0in]{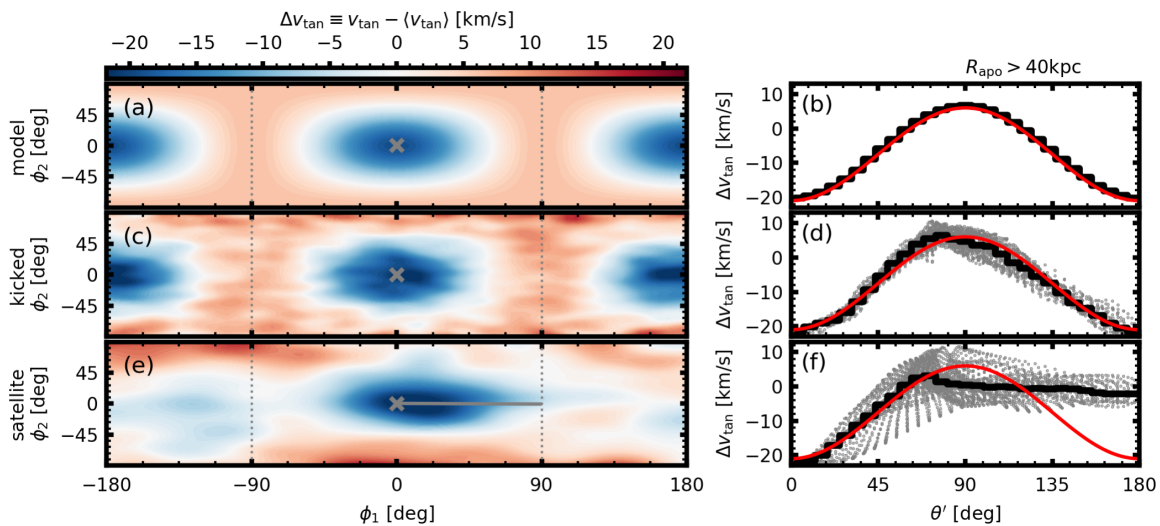} 
\caption{\label{fig:tunneldemo} Panels a,c,e: Tangential velocity deviation ($\dvtan$) reflex motion maps in $\phi_1-\phi_2$ space. Each row corresponds to a different listed model. The present location of the satellite is marked with a gray `x'. In panel e, the trajectory of the satellite to the present location is shown as a gray line along $\phi_2=0$. The map in each panel is computed for stars with apocentres satisfying $\rapo>40$ kpc. Gray dotted lines indicate 90$^\circ$ from the satellite location. Panels b,d,f: Variation in $\dvtan$ radiating outward in angular distance $\tprime$ from the present-day satellite location. The gray points are the discrete points computed for the map. The black curve is the median in each $\tprime$ bin. In each panel, the red curve is a $\sin^2$ fit to the geometric model in panel b.} \end{figure*}

\subsection{Model integration} \label{subsec:bfe}

We evolve the models using {\sc exp}, a basis function expansion (BFE) code \citep{weinberg99,petersen20a}. Briefly, {\sc exp} uses biorthogonal potential-density functions to represent the mean gravitational field and higher-order deviations. We represent the DM halo with a spherical basis, retaining harmonic terms up to $l_{\rm max}=m_{\rm max}=6$ and radial terms up to $n_{\rm max}=32$. We represent the stellar disc with a cylindrical basis composed of two-dimensional functions in radial and vertical dimensions. The disc potential is then represented by only azimuthal harmonics, $m_{\rm max}=6$. We retain radial orders in the disc up to $n_{\rm max}=18$. Example functions are shown in \citet{petersen20a}. The basis is expanded from the centre of energy for each component. To compute the centre of energy, we rank-order the particles as a function of total energy $E=\frac{1}{2}v^2 + \Phi_{\rm tot}$ where $\Phi_{\rm tot}$ is the total potential from the combined disc and DM halo system and average the positions of the 2048 highest-energy particles. We track the centre of energy at each timestep in the simulation, $\Delta T=2.2$ Myr. 

We first run a control simulation with no satellite for 2 Gyr. We run a second simulation with the unbound satellite on a pre-determined trajectory. We create an analytic unbound satellite trajectory in the total MW potential, following numerical modeling that suggests a first-infall scenario for the LMC \citep{besla10}. We integrate the LMC orbit backwards in time in the initial MW potential assuming that the present-day LMC location is pericentre \citep{kallivayalil13, pietrzynski19}, then begin the simulation and follow the satellite along the specified trajectory. The starting position and velocity vectors are $(x,y,z,\dot{x},\dot{y},\dot{z}) = (149.16, 377.82, -299.34, -48.6, -184.8, 94.3)$. We analyse the simulation using a snapshot where the LMC is at $(x,y,z)=(-13.37, -72.89, -24.56)$ kpc with $(\dot{x},\dot{y},\dot{z}) = (-154.6, -163.8, 323.3)$ \kms. Following a predetermined analytic trajectory neglects dynamical friction, tidal deformation, and mass stripping; however, the models of \protect{\citet{gomez15}} suggest that for our parameters in a first-infall scenario, neglecting to model a fully self-consistent LMC over the past Gyr will only minimally bias the model. We define a rotation matrix with angles about the XYZ axes $(\alpha=5^\circ,\beta=45^\circ,\gamma=65^\circ)$ such that the unbound satellite trajectory, originally on a track similar to the LMC, moves along one axis in a newly-defined angular coordinate system $(\phi_1,\phi_2)$. With the rotation, the satellite travels along $\phi_2=0^\circ$ axis and reaches $(\phi_1,\phi_2) = (0^\circ,0^\circ)$ at pericentre passage. The antipode of the satellite location is located at $(\phi_1,\phi_2)=(\pm180^\circ,0^\circ)$.

Figure~\ref{fig:s13motion} shows quantities of interest computed from the unbound satellite model. We use this model to tune the idealised models for comparison. The upper is the separation between the centre of each component (defined by the minimum energy, see Section~\ref{subsec:bfe}) as a function of time in the simulation. The lower panel of Figure~\ref{fig:s13motion} shows the velocity of the components relative to the inertial simulation centre. At the pericentre passage, the disc is moving at $45\kms$ relative to the inertial simulation centre, but only $32\kms$ relative to the DM halo expansion centre.

\section{Isolating Milky Way reflex motion}\label{sec:reflex}

We construct three simplified cases to characterise the reflex motion kinematic signature: 
(1) a theoretical geometric case, where the transformation is analytically computed according to a geometric argument;
(2) an instantaneous kick case, where the disc is displaced at a set velocity;
(3) an unbound infalling satellite case, where the satellite follows a specified trajectory.

{\bf Geometric model.} The theoretical case provides information on the expected morphology of the disc reflex motion. One can imagine the stellar halo as a shell of stars at some fixed distance. We are traveling toward a point $(\phi_1,\phi_2)=(0^\circ,0^\circ)$ with velocity $v_{\rm travel}$. The tangential velocity map holds a particularly rich amount of diagnostic information for determining reflex motion signatures. We refer to the mean-subtracted tangential velocity as the tangential velocity deviation, $\dvtan$. The quantity $\dvtan$ is computed by first calculating the modulus of the tangential velocity\footnote{In the un-rotated frame, $v_{\rm tan} = (v_\phi^2  + v_\theta^2)^{1/2}$.}, $v_{\rm tan} = (v_{\phi_1}^2  + {v_{\phi_2}}^2)^{1/2}$ and then the deviation is defined as $\dvtan\left(\phi_1,\phi_2\right)\equiv v_{\rm tan}(\phi_1,\phi_2) - \langle v_{\rm tan} \rangle$. 

One can imagine the tangential motion on the surface of the sphere. In the direction of travel, the tangential motion will be zero. Perpendicular to the direction of travel, the tangential motion will be $v_{\rm travel}$. Along the plane of travel ($\phi_2=0^\circ$) the functional form is $\sin^2(\phi_1)$. Owing to spherical geometry, perpendicular to the plane of travel, $\phi_1=0^\circ$, the functional form is $\sin^2(\phi_2)$ with an amplitude $v_{\rm travel}$. When $\phi_1,\phi_2 \ne 0^\circ$, the signal is a combination of the velocity along galactic latitude, $v_{\phi_1}\cos (\phi_2)$, and galactic longitude, $v_{\phi_2}$ such that as the distance from the minimum increases, the signal increases as
\begin{equation}
\dvtan^{\rm reflex}\left(\phi_1,\phi_2\right) = v_{\rm travel}\left(\cos^2 \phi_2 \sin^2 \phi_1 + \sin^2 \phi_2 \right).
\label{eq:tunnelsignal}
\end{equation}

We show the model velocity signal in $(\phi_1,\phi_2)$ in panel a of Figure~\ref{fig:tunneldemo}\footnote{Equation~\ref{eq:tunnelsignal} is equivalent to the dipole $l=1,m=\pm1$ spherical harmonic, $Y_{11}(\phi_1,\phi_2)^2$, in line with analytic predictions \protect{\citep{weinberg89}}.}. The present location of the satellite is marked with a gray `x'. We assume that $v_{\rm travel}$ will be approximately the velocity difference between the centre of the disc expansion and the centre of the DM halo expansion at the chosen time, $v_{\rm travel}=32\kms$ (Figure~\ref{fig:s13motion}). In panel b of Figure~\ref{fig:tunneldemo}, we define a new coordinate, $\tprime$, the angular distance between the present satellite location and all points on the spherical distribution. The minimum is at $\tprime=0^\circ$, the maximum is at $\tprime=90^\circ$, and a second minimum is at $\tprime=180^\circ$. The gray points are the individual $(\phi_1,\phi_2)$ points from the full map. The black curves are the median values at each $\tprime$ value. We fit the black curve in red with a $\sin^2$ parameterisation.

{\bf Instantaneous kick model.} After establishing a theoretical geometric foothold, we examine the signal in real models of a spherical halo with a MW-like disc embedded. In this limit, the disc moves relative to the fixed halo. Using a snapshot from the evolved control model, we apply an imposed instantaneous velocity $v_{\rm travel}=32\kms$ (the velocity difference between the centre-of-mass for the disc and DM halo) toward $(\phi_1,\phi_2)=(0^\circ,0^\circ)$ and measure the spherical velocity components. Motivated by observations and a desire to select particles with long dynamical times (see Section~\ref{subsec:apocentre}), we show only particles with $\rapo>40$ kpc, but currently within 20 kpc of the sun. The deviations from the geometric model necessarily come from the distribution of the DM halo, hinting at the possible use of the reflex motion signature to determine the structure of the DM halo. We show the $\dvtan$ map in panel c\footnote{All velocity maps in Figure~\protect{\ref{fig:tunneldemo}} are from the position of the galactic centre, and thus do not have the local solar reflex. All velocity maps are computed using a kernel density estimator.} and the amplitude as a function of $\tprime$ in panel d. The red curve in panel d is the fit from panel b, the geometric model.

{\bf Satellite model}. We show the $\dvtan$ map for the satellite model on a specified trajectory in panel e of Figure~\ref{fig:tunneldemo}. The particles are again filtered by apocentre, as in the previous two models. The trajectory of the satellite is shown as a gray line. The suite of idealised simulations demonstrates that the primary observed structure in the $\dvtan$ map comes from the reflex motion of the MW disc as it moves against the halo stars with long dynamical times. The utility of the transformation into a frame where the satellite is traveling along a single axis is clear: the minimum near $(\phi_1,\phi_2)=(0^\circ,0^\circ)$ is apparent. However, the true minimum is at $\phi_1=15^\circ$, owing to the response of the halo over time lagging behind that of the disc along the satellite trajectory. In de-rotated coordinates, the vector is oriented toward $(\ell,b) = (-78.4^\circ,-47.8^\circ)$. The antipode signal is not located directly at $(\phi_1,\phi_2)=(\pm180^\circ,0^\circ)$ as would be expected from the geometric and kicked models. This owes to the history of the satellite trajectory. The minimum in $\dvtan$ may be used as a diagnostic of the past trajectory of the satellite as well as an indicator of the DM halo structure. We discuss strategies to maximize the antipode signal in Section~\ref{subsec:apocentre}. \citet{garavito19} showed that streaming motions owing to the density wake would produce a distinct velocity pattern in the stellar halo, but did not parameterise the pattern. We also measure wake signals in our model along the LMC trajectory, but we optimize selections of particles to measure the kinematic signature of reflex motion. We address the selection in Section~\ref{subsec:apocentre}. A future work will compare the relative importance of reflex motion and wake signals for realistic samples of stars.

The scatter in the $\dvtan-\tprime$ panels d and f of Figure~\ref{fig:tunneldemo} comes from the non-spherical nature of the reflex motion signal, evident in panels c and e. The deviation from spherical encodes information about the structure of the DM halo; the principal deviation is attributable to adiabatic compression of the DM halo in response to the disc potential (in the case of the kicked model) as well as the trajectory of the satellite along the $\phi_2=0^\circ$ axis in the satellite model. The signature will also depend on the DM halo profile, which we will study in a future contribution.

\begin{figure} \centering \includegraphics[width=2.5in]{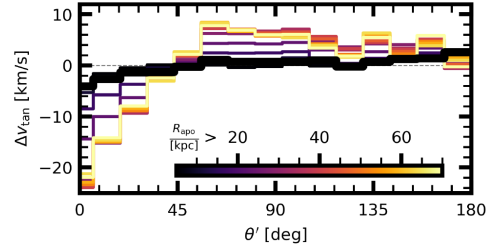} 
\caption{\label{fig:apofilter} Amplitude of the $\dvtan$ signal as a function of $\theta^{\prime}$ for different apocentre limits. The observed stellar halo is defined as stars within 20 kpc of the Sun. The apocentres are defined from the galactic centre. The thick solid black line is the measured amplitude with no apocentre limit applied. } \end{figure}

\section{Discussion} \label{sec:discussion}

We discuss two points important for detecting and interpreting the reflex motion signature: the determination of apocentres for stars to maximise the observed signal (Section~\ref{subsec:apocentre}) and the biases that may complicate detection and/or confuse other samples (Section~\ref{subsec:observationalbiases}).

\subsection{Importance of filtering by apocentre} \label{subsec:apocentre}

With full phase-space information, one may compute the trajectory of the star given some potential model, and hence the apocentre. One may maximise the reflex motion signal in the stellar halo by selecting stars with large apocentres, which have long dynamical times, and thus have not yet had time to respond to the recent infall of the LMC. In our models, we compute the apocentre of each particle by measuring the radius at each timestep in the simulation and recording the largest apocentre from the halo expansion centre. As an observational consideration, we only select halo stars with instantaneous distance $<20$ kpc of the solar location.

In Figure~\ref{fig:apofilter}, we show that choosing stars with larger apocentres results in an increased $\dvtan$ signal: the signal increases in strength as one selects larger apocentres. One minimises $\dvtan$ (maximises the reflex motion signal) at $\tprime=0^\circ$ by selecting stars with $\rapo>40$ kpc. However, the signal at $\tprime>90^\circ$ does not begin to show the decrease to the satellite antipode predicted by the idealised models apart from a selection of stars with $\rapo>70$ kpc. The strength of the signal for large apocentre particles means that selecting tracers at large distances does provide the best constraints on the reflex motion signal, but conversely, will be the most biased by the reflex motion signal if one is attempting to use the tracer as a probe of a static potential. Thus, by using a tracer population at large distances, the bias owing to the reflex motion increases.

\subsection{Observational biases from reflex motion} \label{subsec:observationalbiases}

\begin{figure} \centering \includegraphics[width=3.45in]{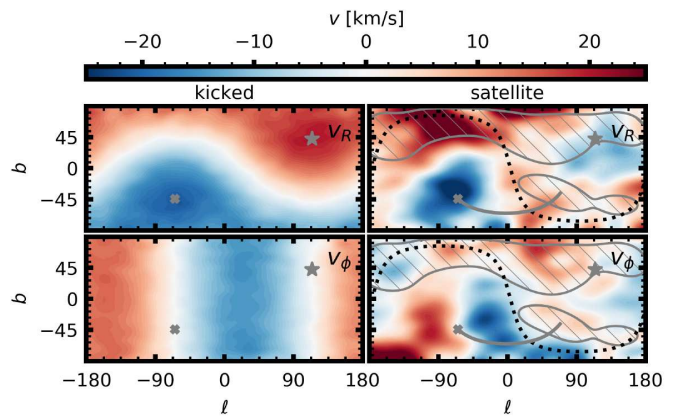}
\caption{\label{fig:coveragemap} Galactic ($\ell,b$) maps of radial velocity $v_R$ (upper panels) and azimuthal velocity $v_\phi$ (lower panels). The left column is the kicked model; the right column is the satellite model. We show the satellite trajectory and current position as a gray line and gray `x'. The satellite antipode is marked with a gray $\star$. The orbital plane of the Sagittarius dwarf galaxy is shown as the black dotted line \protect{\citep{law10}}. The approximate coverage of SEGUE spectra is shown as a gray hatched region.} \end{figure}

In the right column of Figure~\ref{fig:coveragemap} we demonstrate potentials for observational bias on radial and azimuthal velocity maps. Given the large-scale velocity variations inherent to the reflex motion signal, one must carefully consider whether the coverage on the sky is sufficient to observe all-sky trends. We show the approximate coverage of BHB stars from SEGUE \citep{xue11} used by \citet{deason17} to determine $\langle v_\phi \rangle$ in the stellar halo in Figure~\ref{fig:coveragemap} as a gray hatched region. SEGUE \citep{yanny09} covered over 14000 square degrees: the coverage may be sufficient to recover the large-scale signal. In the left column of Figure~\ref{fig:coveragemap} we demonstrate the de-rotated radial and azimuthal velocity maps in the kicked model. Comparison between the kicked and satellite models suggest that the bulk of the all-sky velocity signal in nearby stars with large apocentres comes from reflex motion, rather than the wake.

We also show the orbital plane of the Sagittarius dwarf computed from \citet{law10} in the right panels of Figure~\ref{fig:coveragemap}. Unfortunately, Sagittarius is located near $\tprime=90^\circ$ relative to the present-day location of the LMC, which possibly confuses the signal of the reflex motion and necessitates the modeling of the Sagittarius stream and the LMC infall simultaneously\footnote{With a present-day mass of $M_{\rm Sgr}=5\times10^8\msun$ \protect{\citep{niederste10}}, Sagittarius will not strongly displace the MW barycentre itself.}. The leading arm of Sagittarius is particularly subject to the reflex motion of the MW, contributing up to $v_R=30\kms$ to the measured line-of-sight signal. Such bias calls for caution when fitting orbits to tidal streams, as one may bias the potential if reflex motion velocities are not taken into account. This may be the case for long tidal streams such as Orphan and GD-1, which span tens of degrees on the sky. Solid-body barycentric motion is insufficient to resolve the reflex motion bias owing to the differential effect in the halo with apocentres. While radial velocities are expensive to observe,  all-sky coverage of a large sample of tracers and a suite of self-consistent N-body models is necessary to disentangle reflex motion from other dynamical effects (e.g. rotation, substructure or the density wake caused the LMC infall). We identify three important signposts in the reflex motion signature: (i) The $\dvtan$ signature in the outer stellar halo, maximised by detecting particles with large $\rapo$. (ii) The $v_r$ and $\dvtan$ signature near the satellite itself, which is not dominated by the local effect but instead the reflex motion in our satellite model. (iii) The radius where the halo stops being dragged with the inner halo. The inner halo responds as the disc, with the stellar halo gradually becoming independent of the disc motion with increasing apocentres. Placing observational constraints on each of the signposts will inform future models of the reflex motion and time-dependent models for the MW system.

\section{Conclusion} \label{sec:conclusion}

We present a model for the MW-LMC system and identify signatures of reflex motion of the stellar disc owing to the presence of the satellite moving the disc barycentre. We demonstrate that reflex motion creates a measurable global signal in the stellar halo of the MW. We identify key locations in velocity maps: the direction of travel of the disc, the antipode of the disc travel direction, and the maximum of the tangential velocity signal. We measure the strongest signals from stars that sample the outer halo potential: those with large apocentres. To constrain apocentres for individual stars, we require a six-dimensional data set and the model MW potential.

We construct idealised models to explain the observed reflex motion signature, using maps of the tangential velocity as the crucial metric with which to determine the imprint of reflex motion. We build an analytic dipole model using simple geometric arguments that may be applied to the real MW-LMC system to test for signatures of reflex motion using existing and future data sets. We find that simple barycentric motion of the DM halo is insufficient to reproduce the all-sky effects observed in the model: one needs self-consistent $N$-body models, not only a rigid centre-of-mass motion, to make MW potential measurements using stellar halo tracers. If the global patterns are not taken into account, one could bias radial and tangential velocity measurements by up to the disc travel velocity, which may be as high as 40$\kms$ for an LMC with $\mlmc=10^{11}\msun$ \citep{gomez15}. The bias in measurements is largest for the outer tracers of the potential, those with long dynamical times -- and different tracers may show different signatures depending on their $\rapo$ distributions. Given future radial velocity measurements and more sophisticated models for the MW, we will be able to constrain the structure of the DM halo through comparison with reflex motion $N$-body models. 

\citet{erkal19} predicted a bulk upward motion in the stellar halo. Our models indicate that the realistic signal will be more complex than a bulk upward motion; the motion will be perpendicular to the trajectory of the LMC and vary with distance.

Two major future lines of study will assist reflex motion interpretation. First, the DM halo density profile, the shape, and the concentration of the DM halo are not well-constrained in the MW, nor well-studied in models that include the LMC. The structure of the DM halo will be the primary factor controlling the reflex motion signatures. Second, many LMC quantities are not well constrained: the exact trajectory and mass of the satellite are crucial ingredients to isolate the reflex motion signatures. In particular, if the LMC is not infalling for the first time, the signature of reflex motion will still be present, but may be appreciably different. Our LMC model, a rigid softened point source, will not capture the full dynamics of the system. The inclusion of a self-consistent LMC may incite a more complex time-dependent MW potential due to the stripping of the LMC, including the capacity for increased reflex motion if the pericentre passage is longer in duration. Both avenues deserves focused study to determine the effect of including or excluding specific ingredients on the reflex motion models.

\section*{acknowledgements}

We thank the anonymous referee for comments that strengthened the work. MSP thanks Martin Weinberg for helpful discussions. This project made use of {\it numpy} \citep{numpy} and {\it matplotlib} \citep{matplotlib}.

\bibliography{ReflexMotion}
\label{lastpage}

\end{document}